# Gain Induced Topological Response via Tailored Long-range Interactions

Yuzhou G. N. Liu[1,2], Pawel S. Jung[2,3], Midya Parto[2],

Demetrios N. Christodoulides[2,*], Mercedeh Khajavikhan[1,2,*]

[1]Ming Hsieh Department of Electrical and Computer Engineering, University of Southern California, Los Angeles, California 90089, USA

[2]CREOL, The College of Optics & Photonics, University of Central Florida, Orlando, Florida 32816–2700, USA

[3]Faculty of Physics, Warsaw University of Technology, Koszykowa 75, 00-662 Warsaw, Poland

*Corresponding authors: demetri@creol.ucf.edu (D.N.C) & khajavik@usc.edu (M.K.)

**The ability to tailor the hopping interactions between the constituent elements of a physical system could enable the observation of unusual phenomena that are otherwise inaccessible in standard settings[1,2]. In this regard, a number of recent theoretical studies have indicated that an asymmetry in either the short- or long-range complex exchange constants can lead to counterintuitive effects, for example, the possibility of a Kramer's degeneracy even in the absence of spin 1/2 or the breakdown of the bulk-boundary correspondence[3-8]. Here, we show how such tailored asymmetric interactions can be realized in photonic integrated platforms by exploiting non-Hermitian concepts, enabling a class of topological behaviors induced by optical gain. As a demonstration, we implement the Haldane model, a canonical lattice that relies on asymmetric long-range hopping in order to exhibit quantum Hall behavior without a net external magnetic flux. The topological response observed in this lattice is a result of gain and vanishes in a passive but otherwise identical structure. Our findings not only enable the realization of a wide class of non-trivial phenomena associated with tailored interactions, but also opens up avenues to study the role of gain and nonlinearity in topological systems in the presence of quantum noise.**

The interplay between long- and short-range interactions is known to play a pivotal role in many and diverse areas of physical sciences. Such interactions naturally manifest themselves in Rydberg atoms and ionic systems and are known to drastically modify the metal-insulator transitions in Bose-Hubbard models[2,9]. In the realm of topological physics, the possibility of introducing asymmetries in the complex exchange constants of a system ($J_{ij} \neq J_{ji}$) has been actively pursued along several theoretical fronts in an effort to unravel novel effects and behaviors. In this respect, a host of intriguing topological phenomena has been predicted, ranging from the quantum anomalous Hall and non-Hermitian skin effects, to Anderson transitions in the Hatano-Nelson model and the emergence of Gaussian symplectic ensembles in the absence of spin[3–8].

An archetypical lattice whereby a topological phase can appear because of non-symmetric long-range exchange interactions is that proposed by Haldane in 1988[1]. This proposition trailblazed a path towards the discovery of topological insulator materials based on alternative processes like for example spin-orbit interactions[10–17]. Yet, despite a series of groundbreaking advancements, to this day, the original Haldane lattice, has remained an elusive crystal within the context of solid-state physics, yet to be synthesized in the lab. What makes the implementation of this system and other related topological models challenging, is the difficultly of realizing the aforementioned non-symmetric long-range exchange mechanisms in condensed matter. Clearly of interest will be to explore alternative platforms where this type of interactions can be supported[18].

Photonics has so far provided a versatile testbed to study a variety of topological effects through artificial gauge fields and synthetic dimensions[19–35]. However, most interactions, in current photonic topological lattices tend to be of the nearest neighbor type and symmetric. Ideally, it will be beneficial if one had the freedom to establish any arbitrary interconnectivity, irrespective of whether it is symmetric/asymmetric or short/long range. In this work, we show how optical gain and non-Hermiticity can be utilized to one's advantage in order to tailor the hopping long-range exchange between resonant elements in a photonic arrangement[36]. We demonstrate this approach by implementing a Haldane lattice operating in its topological Chern regime. As we will show, the topological features of this lattice are manifested solely because of non-Hermiticity and nonlinearity. In this respect, we showcase the potential of active photonic systems in designing a class of topological lattices that are uniquely enabled by optical gain.

The Haldane structure[1] features a honeycomb lattice, comprised of two species of atoms (Fig. 1a). Besides nearest neighbor (NN) interactions among detuned neighboring atoms (that break the inversion symmetry), an asymmetric exchange is established between the next-nearest neighboring (NNN) elements that can be expressed through antisymmetric complex hopping terms of the form $t_2 e^{\pm i\varphi}$. The Hamiltonian associated with this lattice is expressed as follows:

$$\widehat{H} = M \sum_{p \in A} \hat{a}_p^\dagger \hat{a}_p - M \sum_{p \in B} \hat{a}_p^\dagger \hat{a}_p + t_1 \sum_{NN} \hat{a}_p^\dagger \hat{a}_q + t_2 \sum_{NNN} e^{i\varphi_{pq}} \hat{a}_p^\dagger \hat{a}_q \quad (1)$$

where $\hat{a}_p^\dagger$ and $\hat{a}_q$ are the bosonic creation and annihilation operators at sites $p$ and $q$, respectively. $M$ represents a relative detuning (energy difference) associated with the two sublattice sites $A$ and $B$, $t_1$ is the coupling strength between neighboring elements, while $t_2$ stands for the next nearest neighbor hopping coefficient, with $\varphi_{pq} = -\varphi_{qp}$ being the phase associated with the NNN exchange. Both $t_1$ and $t_2$ are taken here to be real and positive. A characteristic feature of the Haldane two-band model is a phase transition between trivial and topological states, depending on the complex tunneling amplitude among next-nearest neighboring sites. Here, by varying the values of $M, t_1$ and $t_2$, the Haldane lattice can transition from a topologically trivial to a non-trivial phase. When terminated with a proper arrangement, for example with a zigzag edge, this topological structure, can support a unidirectional and scatter free edge current. Figures 1b-d display the topological and trivial regimes arising at different values of the parameters $M$, $t_2$ and $\varphi$ (when $t_1 = 1$). As evident from this figure, when $t_2 = 0$, the array is in a topologically trivial phase with a Chern number $C = 0$. On the other hand, by tuning the NNN hopping amplitude to $M/t_2 = 1.25$ and $\varphi = \pi/2$, this system will behave in the Chern-insulating regime[1] ($C = +1$).

In our study, we realize a topological Haldane lattice by using an active photonic structure as shown in Fig. 2. Its unit cell, depicted in the inset, is comprised of two sets of ring-type cavities, A and B (color coded by red and blue). These two species of resonators have slightly different perimeters, thus supporting eigenfrequencies that are somewhat detuned. A fan-shaped construct is incorporated in each active resonator that generates an exceptional point and enforces unidirectional flow of light due to the interplay of gain, spontaneous emission, and gain induced nonlinearity with the geometrical features of the cavity (Supplementary Sections 1, 2)[28,37]. The use of an exceptional point guarantees that the ring remains single moded, thus supporting propagation in only one direction (Supplementary Section 3)[37]. This unique property, a direct consequence of non-Hermiticity, is of central importance in realizing the Haldane lattice as it enables the antisymmetric coupling. Here, the neighboring elements are designed to support counter-propagating modes, allowing light in adjacent resonators to evanescently couple to each other through a coupling strength $t_1$ that is set by controlling the distance between the two bordering resonators. Even though the constituent elements of this lattice are by nature non-Hermitian and nonlinear, their interactions are governed by the Hermitian Hamiltonian of Eq. (1).

As previously indicated, the most crucial feature of this lattice, is the complex antisymmetric long-range NNN hopping between elements of the same species[1] (A↔A and B↔B). Typically, in passive optical arrangements, the Haldane model requires time-reversal symmetry breaking, which is difficult to achieve without the use of magneto-optic materials that rely on external magnetic fields. Here we propose an innovative approach in order to address this issue. In this respect, we couple two unidirectional active cavities (of the same type) through a combination of directional couplers and waveguides, having a total length of $L$ (Methods). Given that all NNN resonators allow light to circulate in the same direction (here for example counter-clockwise), light that couples from resonator ① to ② (inset in Fig. 2a) has to traverse the "through" ports of the two couplers, thus acquiring an overall phase shift of $\phi = kL$ (see Supplementary Section 4), where $k$ is the wave vector in the waveguides involved. On the other hand, for light traveling from resonator ② to ①, it has to follow the "cross" ports path, resulting in an additional $\pi$ phase shift (i.e. the overall phase is now $\pi + \phi$). In either direction, the magnitude of the complex NNN hopping coefficient has the same value $t_2$. By properly choosing the length of the link ($L$) so as $\phi = \pi$, the hopping coefficients assume the values $it_2$ and $-it_2$ when considering the paths ① → ② and ② → ①, respectively. Since in this case, $\varphi_{pq} = \pi/2$, the system is placed in the Chern insulating regime ($C = +1$) as depicted in Fig. 1b. This complex antisymmetric coupling between NNN elements of the same species is what actually enables the Haldane model to acquire a topological phase. Finally, in order to provide coupling between all the NNN resonators in the hexagonal unit cell, it is imperative to allow the waveguide sections to intersect each other. For high contrast waveguides (InGaAsP-SiO$_2$/air as used here), electromagnetic simulations confirm that the crosstalk between intersecting channels is negligible provided that the crossing angle is approximately 90 degrees[38,39] (see Supplementary Section 5). More details about the design have been provided in Methods.

In our work, the Haldane lattice was fabricated on a wafer with InGaAsP multiple quantum wells. The detailed fabrication process is described in the Methods. The lattice has been terminated in zigzag edges, naturally resulting in an equilateral triangular structure. It should be noted that other edges like armchair can also be realized, resulting in ribbon-like lattices (see Supplementary Section 6). Overall, this system contains 166 micro-resonators (or 66 unit cells). In addition, we incorporated three gratings at the corners of the structure in order to monitor the light exiting the lattice at these points. SEM and microscope images of our photonic Haldane lattice are shown in Figs. 2b-d. The fabricated samples are then characterized in a photoluminescence measurement station, where an optical mask placed in the path of the pump beam to generate the desired pump profile on the sample. In order to promote the topological edge mode, the outer perimeter of this Haldane lattice is optically pumped with a pulsed laser. The emission profile is then captured by an InGaAs infrared camera, while the lasing spectrum is collected with a monochromator accompanied by an InGaAs detector array. The sample was placed in a cryostat with an electric heater in order to fine-tune the optical length ($L$) of the waveguide sections, in order to induce the topological phase (Supplementary Section 7). The details of the characterization station is explained in the Methods.

In our experiments, we first characterize the complex antisymmetric hopping behavior in a constituent sub-unit comprising of two identical unidirectional resonators as shown in Fig. 3a. The two grating ports are incorporated to probe the emission from the structure at a wavelength $\lambda \approx 1550$ nm. This coupled system supports two eigenmodes $[1 \quad \pm i]^T$ with corresponding eigenvalues of $\pm t_2 e^{i\phi}$ (Supplementary Section 4). It should be noted that, if the two involved resonators are not enforced to be unidirectional, an additional coupling will form between the counter propagating modes in these two resonators which can prevent the system from displaying a topological response (Supplementary Section 8). When this structure is pumped, depending on the value of $\phi$, various emission profiles and spectra are observed. If for example, $\phi = \pi/2$ or $3\pi/2$, the two eigenvalues become purely imaginary (with opposite signs), hence one of the two modes experiences a substantially higher gain. Consequently, the emission spectra of the corresponding lasers become single-moded, and the output emissions from the two resonators constructively interfere, leading to a power build-up in one of the two gratings (see Figs. 3b and d). On the other hand, when $\phi = 0$ or $\pi$, the two eigenvalues are real and therefore one should observe the simultaneous lasing of two modes with a frequency splitting of $2t_2$. As a result, the power is equally distributed between the two grating ports (see Fig. 3c). As indicated earlier, this is the phase condition required for the NNN hopping (generally $\phi = n\pi$ where $n$ is an integer). Similarly, we examined the behavior of the 3-element sub unit-cell (comprised of identical triangular resonators). Here, when one of the three elements is pumped, the energy is expected to flow in one of the two counter-propagating directions around the sub unit cell (Figs. 3f-l). As shown in Figs. 3j-l, for $\phi$s that are centered at odd values of $n$, the mode circulates in the clockwise direction, while for $\phi$s positioned around the even values of $n$ it selects the counter-clockwise path. In this regard, when integrating these sub-units in the greater Haldane lattice, the direction of the flow of light can be flipped by either altering the chirality of all the individual resonators involved (by reversing the fan-shaped structures) or by changing the value of $n$ by an odd integer increment.

In characterizing the topological Haldane lattice, we look for its most prominent signature, i.e. a unidirectional flow of light around the perimeter when the array is terminated with zigzag edges. Here, a metallic mask is used to selectively pump the periphery, while an additional attenuating filter is used to cover one of the sides of the lattice from the pump beam (for details about the metallic mask fabrication, see Methods). In order to observe the unidirectional flow of light around the lattice, one edge of the lattice has been kept almost unpumped, while the other two edges are fully pumped. Under these conditions, we expect to observe a power difference between the ports located at the two ends the unpumped edge because of the unidirectional flow of light around the perimeter (see Supplementary Section 9). By monitoring and comparing the intensity of light exiting the grating couplers (incorporated at the corners of the structure), the direction of light transport around the edges of the array is determined. Figure 4 shows the related experimental results. In particular, a comparison between the light intensity levels as they exit the gratings at the two ends of the unpumped side reveals that in the topological Haldane lattice, light circulates in one of the two counter propagating directions (in this case: counter-clockwise), as expected from a Haldane structure. This chiral behavior is consistently observed as we rotate the unpumped edge (see Figs. 4f and g). These results are in full agreement with our theoretical predictions and simulations (see Supplementary Section 9). It should be noted that because the output intensity at the grating couplers is a combination of stimulated and spontaneous emissions, the power ratios reported in the caption of Figs. 4e-g assume finite values. Here, we intentionally chose a structure in which the direction of energy flow around the edge differs from that of the power circulation in the individual resonators involved. The corresponding bandgap of the Haldane topological lattice is calculated to be 200 GHz (see Supplementary Section 10). In order to compare with the trivial case, we fabricated a non-topological lattice by placing the link farther apart from the resonators, thus reducing the strength of NNN hopping. The reported unidirectional power flow is only observed in topological lattices (where $C = 1$) while in a trivial Haldane lattice, light shows no preferential direction when circulating around the array, i.e., the intensity emitted by the two gratings is almost the same (Figs. 4i-k). Finally, in order for this structure to support a chiral edge mode, the directionality of the complex conjugate NNN hoppings for both sets of resonators (A↔A and B↔B) must be identical. Given that this condition has to be satisfied while $\phi = n\pi$ is reached, it limits the number of longitudinal modes that can lase in a topological edge state. In this regard, above threshold, the topological Haldane lattice in expected to emit as a single-moded device. This behavior is observed in our experiments, as is apparent from the emission spectrum provided in Fig. 4h. In contrast, the trivial lattice, having no constraint of this kind, tends to display a multi-mode lasing behavior (Fig. 4l).

In conclusion, we have shown how tailored short- and long-range interactions can be realized in active optical platforms using some of the unique properties offered by gain materials. By adopting this approach, we have demonstrated a topological Haldane lattice operating in its Chern regime. Our work may inspire further developments in the field of topological physics by introducing higher degrees of tailored interconnectivity in Haldane-like networks that could display previously unattainable topological phases.

# Methods

**Analysis.** In this part, we show the emergence of imaginary but Hermitian coupling in judiciously designed active photonic molecules. The schematic of a two-element system with a delayed coupling is depicted in Extended Data Fig. 1a. This type of molecules has been used in our Haldane lattice to realize the desired next-nearest-neighbor (NNN) interactions. The structure is composed of a pair of active (displaying gain) ring resonators with internal S-bends, connected to each other through a combination of directional couplers that are separated by a link waveguide. The S-bends are designed in such a way to enforce unidirectional lasing[40] in the rings (counterclockwise). The left (right) ports of the first (second) directional coupler are weakly coupled to the opposing sides of ring ②(①).

In our analysis, we consider the interaction between the counterclockwise fields in the two cavities (dictated by the direction of the S-bends). In order to keep the number of variables limited, we study the case where all coupling strengths between resonators and waveguides are equal and the two directional couplers are identical. In principle, each ring, when uncoupled, can support a number of spectral lines. Without loss of generality, here we limit our analysis to a single longitudinal mode, since no energy exchange is expected between them. We also assume that the waveguiding section is designed so as to primarily support $TE_0$ transverse mode. In this respect, the interplay between the electric modal fields in the two identical rings can be effectively described through a set of time-dependent coupled equations:

$$\begin{cases} i\dot{a} + \omega_0 a + \kappa_{2\to1} b = 0 \\ i\dot{b} + \omega_0 b + \kappa_{1\to2} a = 0 \end{cases} \quad (2)$$

where $a$ and $b$ represent the modal amplitudes in the two cavities ① and ②, respectively. The angular frequency $\omega_0$ is determined by the resonance conditions for each resonator in the absence of coupling. Using either the temporal or spatial coupled mode analysis, one can readily show that the coupling from resonator ① → ② is given by $\kappa_{2\to1} = i\kappa^2 r^2 e^{ikL}$, while that from ② → ① is expressed by $\kappa_{1\to2} = -i\kappa^2 t^2 e^{ikL}$. Here, $r$ and $t$ represent the through and cross coupling terms for a directional coupler. The coupling strengths between the rings and bus waveguides are all set to be $\kappa$ and the effective propagation constant is $k = 2\pi/\lambda$. Finally, the overall length of the delayed coupling part is $L = L_m + 2L_c$, where $L_m$ and $L_c$ are the lengths of the waveguide sections between the two directional couplers and from each directional coupler to the adjacent ring, respectively. In the case of $r = t$ (3dB directional couplers), when $kL = n\pi$, the two coupling coefficients become $\kappa_{1\to2} = -i\kappa^2/2$ and $\kappa_{2\to1} = +i\kappa^2/2$. Consequently, this system exhibits asymmetric imaginary (yet Hermitian) hopping coefficients that are required to realize the Haldane Hamiltonian.

**Electromagnetic/Wafer Design.** The dimensions of the resonators, the fan-shaped couplers and the links, are shown in Extended Data Fig. 2a-c. In order to realize the required detuning between the two species of resonators, the perimeters of the two rings are designed to be slightly different ($L_A = 62.252$ μm, $L_B = 62.254$ μm). The gain material used in this study consists of six quantum wells of $In_{x=0.56}Ga_{1-x}As_{y=0.938}P_{1-y}$ (10 nm thick)/$In_{x=0.734}Ga_{1-x}As_{y=0.57}P_{1-y}$ (20 nm thick), with an overall height of 200 nm grown on p-type InP substrate[41]. The quantum wells are covered by a 10 nm thick InP over-layer for protection. The optical waveguiding sections are designed to have a width of 500 nm and a thickness of 200 nm, and are partially embedded in a silicon dioxide ($SiO_2$) cladding, so as to operate in the fundamental transverse electric mode ($TE_0$) at a wavelength of $\lambda \approx 1550$ nm. This structure is expected to support the $TE_0$ mode (Extended Data Fig. 2d) with an effective index of $n_{eff} = 2.272$ and a group index of $n_g \approx 4$.

**Fabrication.** The fabrication steps that followed in order to realize the proposed lattices are shown in Extended Data Fig. 3. Here, an XR-1541 hydrogen silsesquioxane (HSQ) solution in methyl isobutyl ketone (MIBK) is used as a negative electron beam resist. The resist is spun onto the wafer (thickness ≈ 50 nm), and is soft baked at a temperature of 180°C (Extended Data Fig. 3a). The rings are then patterned by electron beam lithography (Extended Data Fig. 3b). The wafer is next immersed in tetramethylammonium hydroxide (TMAH) for 120 seconds to develop the patterns and then is rinsed in isopropyl alcohol (IPA) for 30 seconds. The HSQ that is exposed by the electron beam remains and serves as a mask for the subsequent reactive ion etching processes. To perform the dry etching, a mixture of H2:CH4:Ar gases is used with a ratio of 40: 10: 7 sccm. Both RF and ICP powers were set at 150 W and the chamber pressure was held at 35 mT (Extended Data Fig. 3c). The wafer is then cleaned with oxygen plasma to remove organic contaminations and polymers that form during the dry etching process (O2: 50 sccm flow, RIE power: 150 W, ICP power: 150 W, chamber pressure: 50 mT). The patterns are then submerged in buffered oxide etch (BOE) for 10 seconds to remove the HSQ mask (Extended Data Fig. 3d). After this, a 2 μm layer of $SiO_2$ is deposited onto the wafer using plasma-enhanced chemical vapor deposition (PECVD) (Extended Data Fig. 3e). We used SU-8 3010 photoresist to bond the wafer to a glass substrate for mechanical support (Extended Data Fig. 3f). After spinning the photoresist, the sample is placed on the glass, pattern-side-down, and exposed for 30 seconds on both sides. Lastly, the remaining InP substrate is entirely removed by wet etching in hydrochloric acid (HCl) for 100 min (Extended Data Fig. 3g)[42].

In the experiments, a metallic mask is placed in the pump branch to create the desired pump profile on the sample (see Characterization setup section). This mask is fabricated by depositing metals on a glass slide. A pattern is first transferred on the glass slide using photolithography (negative photoresist: NR7-3000, developer: RD6). A layer of gold with a thickness of 10 nm is deposited (for adhesion purposes) using e-beam evaporation, followed by a layer of titanium with a thickness of 100 nm that ensures the transmittance of the pump beam ($\lambda = 1064$ nm) remains below 1%. Finally, the metal is removed from the undesired places by dissolving its underlying resist in acetone.

**Characterization setup.** A micro-photoluminescence (μ-PL) setup, depicted in Extended Data Fig. 4, is used to characterize the structures. The sample is optically pumped by a pulsed (duration: 15 ns, repetition rate: 290 kHz) laser operating at a wavelength of 1064 nm (SPI fiber laser). A beam shaping system with additional metal mask and knife edges is designed to realize the desired pump size/shape. A 10 × microscope objective (NA: 0.26) is used to project the pump beam on the structure and also to collect the photoluminescent emission from the sample. For temperature tuning, the sample is inserted into a cryostat (Janis ST-500). The surface of the sample is imaged by two cascaded 4-f imaging systems into an IR camera (Xenics Inc.). A broadband amplified spontaneous emission (ASE) device is used to illuminate the sample in order to properly position the pump beam with respect to the pattern. A notch filter is placed in the path of emission to attenuate the pump beam. The output spectra are obtained by a monochromator equipped with an attached linear array InGaAs detector. A power-meter is inserted at the focus of the emitted beam in order to measure the output power.

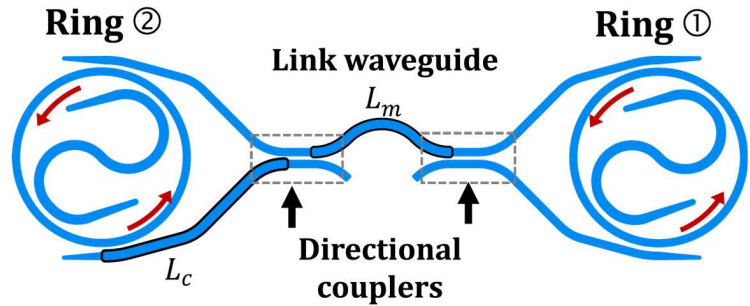

**Extended Data Fig. 1 | Schematic of a two-element system with unidirectional microring resonators and a link structure.** The directional couplers provide a $\pi$ phase difference between the coupling terms $\kappa_{1\to 2}$ and $\kappa_{2\to 1}$. The coupling phase $kL$ can be changed by varying the length $L = 2L_c + L_m$.

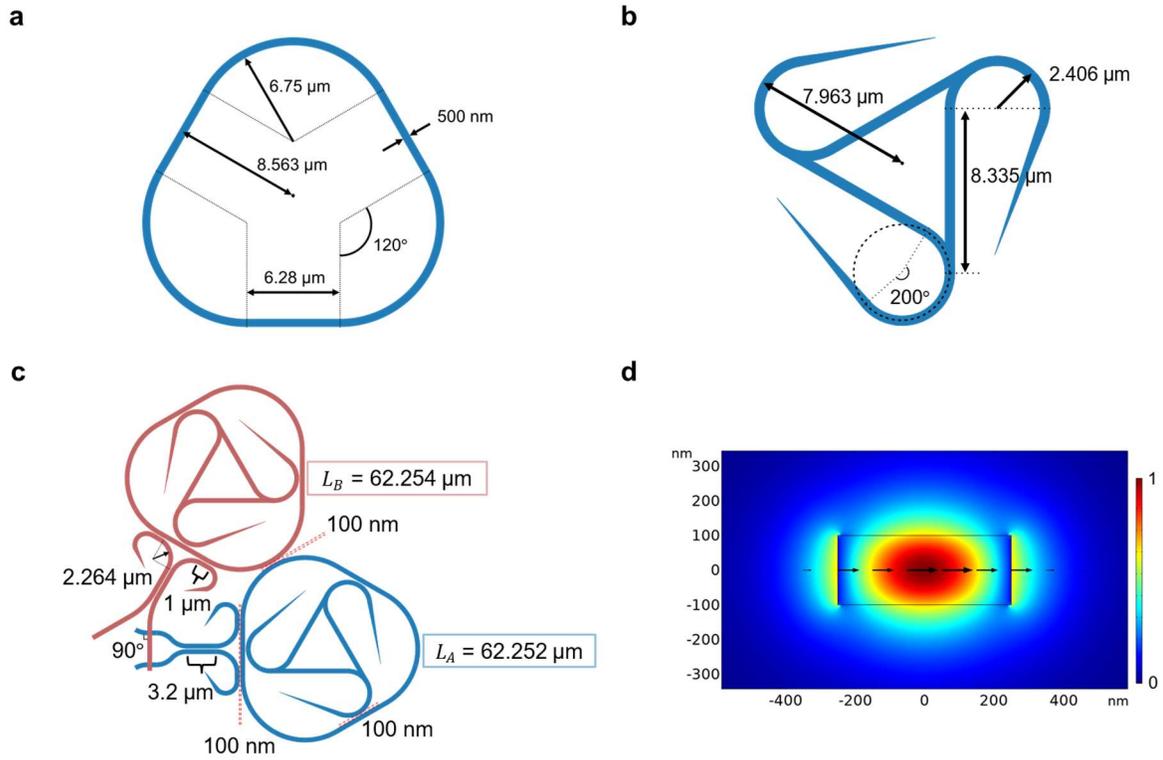

**Extended Data Fig. 2 | Schematics and dimensions of the waveguides, resonators, fan-shape constructs and the link. a,** The triangular microring resonator. **b,** The fan-shaped structure to be incorporated in the triangular resonators. **c,** The perimeters of two adjacent resonators are denoted by $L_A$ and $L_B$. **d,** Transverse electric field distribution in a single waveguide. The black arrows indicate the electric field vector.

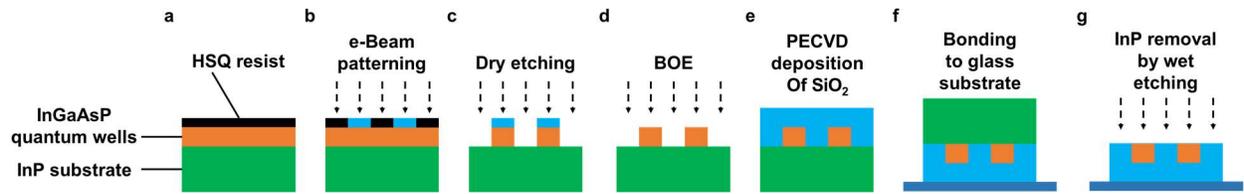

**Extended Data Fig. 3 | Schematic of the fabrication procedure of microring lasers. a,** HSQ e-beam resist is spun onto the wafer. **b,** The wafer is patterned by e-beam lithography. **c,** A dry etching process to define the rings. **d,** The sample is immersed in BOE to remove the masking HSQ. **e,** A 2 $\mu m$ layer of SiO$_2$ is deposited via PECVD. **f,** The wafer is flipped upside-down and bonded to a glass substrate by SU-8 photoresist to provide mechanical support. **g,** Lastly, the InP substrate is wet etched by HCl.

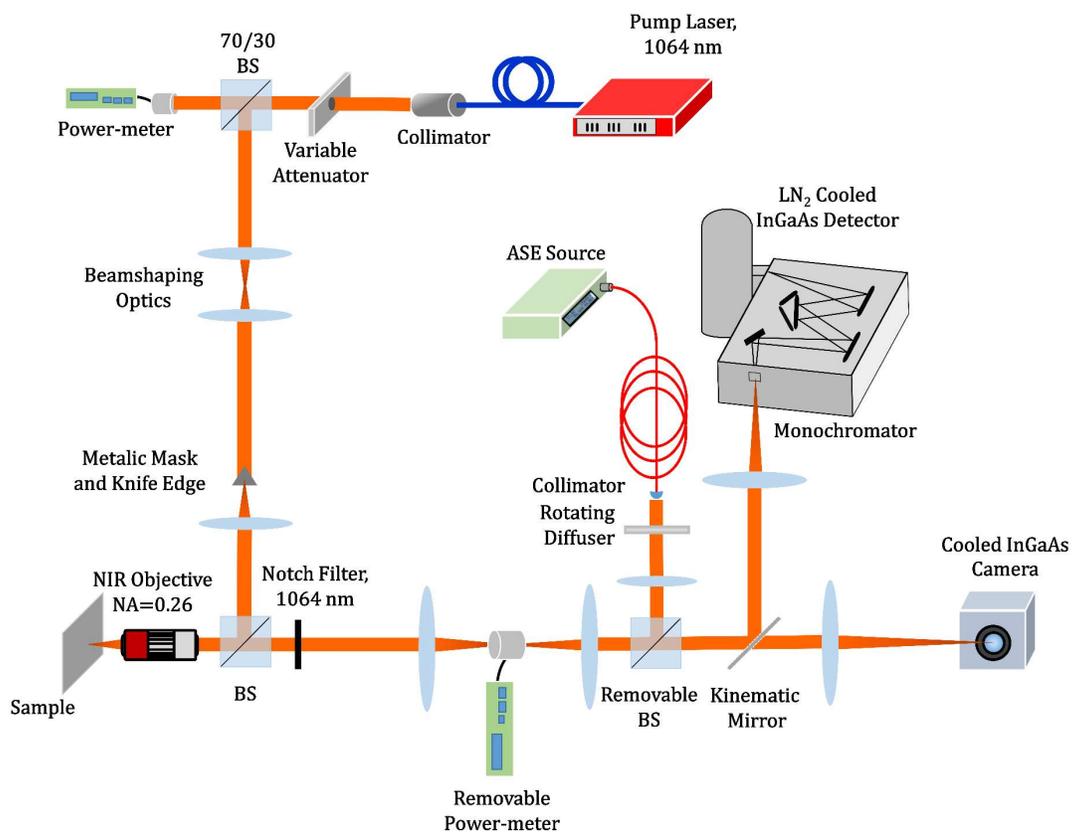

**Extended Data Fig. 4 | Schematic of the μ-PL characterization setup.** The microrings are pumped by a pulsed laser (15 ns pulse width, 290 kHz repetition rate). The pump beam is focused onto the sample with a 10 × objective, this objective in turns also collects the emission from the samples. Light is then either directed to a linear array detector for spectral measurements or to an IR camera for intensity profile observation.

**Acknowledgements**
This work was supported by the DARPA (D18AP00058), Office of Naval Research (N00014-20-1-2522, N00014-20-1-2789, N00014-16-1-2640, N00014-18-1-2347, N00014-19-1-2052), Army Research Office (W911NF-17-1-0481), Air Force Office of Scientific Research (FA9550-14-1-0037, FA9550-20-1-0322), National Science Foundation (CBET 1805200, ECCS 2000538, ECCS 2011171), US–Israel Binational Science Foundation (BSF; 2016381), and Polish Ministry of Science and Higher Education (Mobility Plus, 1654/MOB/V/2017).

**Author contributions**
D.N.C. and M.K. conceived the idea. Y.G.N.L designed the structures and experiments. Y.G.N.L fabricated and characterized the lattices and sub-lattices. Y.G.N.L, P.J. and M.P. performed the simulations. Y.G.N.L, P.J., M.P., D.N.C and M.K. developed the theoretical analysis. All authors contributed in preparing the manuscript.

**Competing interests**
The authors declare no competing interests.

**Supplementary Information** is available for this paper.

**Additional information**
**Reprints and permissions information** is available at www.nature.com/reprints.

**Correspondence and requests for materials** should be addressed to
demetri@creol.ucf.edu
khajavik@usc.edu

**Data availability**
Source data are available for this paper. All other data that support the plots within this paper and other findings of this study are available from the corresponding author upon reasonable request.

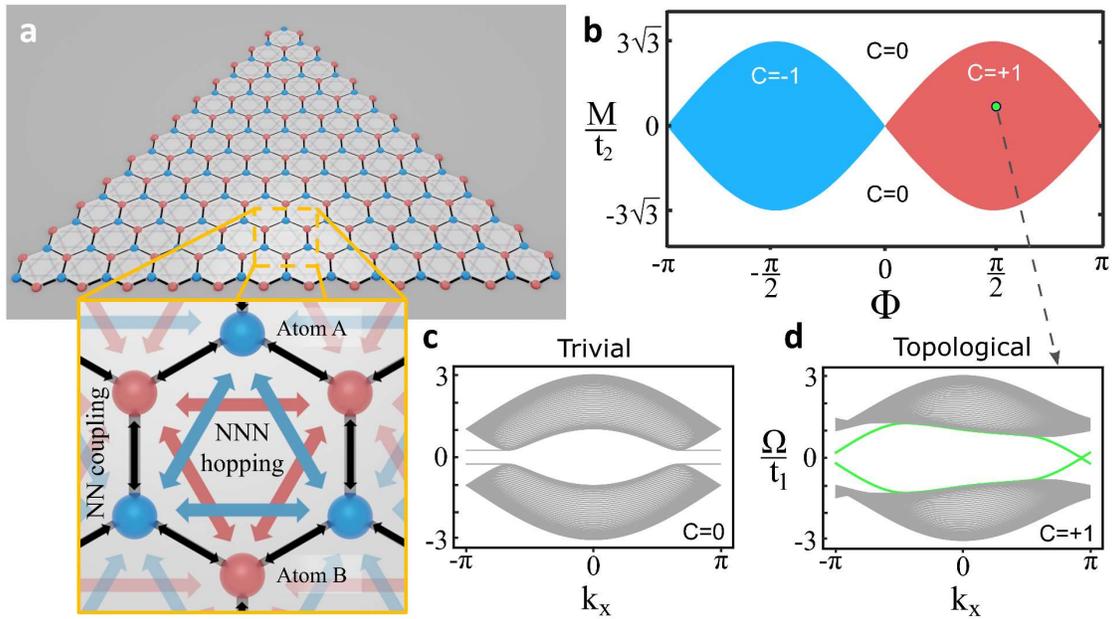

**Fig. 1 | Haldane crystal model. a,** A Haldane lattice featuring two species of atoms (A and B) positioned in a honeycomb arrangement. The inset shows the unit-cell. The nearest-neighbor exchange (black arrows) is symmetric, while the next-nearest neighbor hoppings (red and blue arrows) are asymmetric. **b,** Chern number regimes associated with the Haldane model. The trivial regime has a Chern number $C = 0$ while the topological regimes display Chern numbers $C = \pm 1$. **c,** The band structure in the trivial ($C = 0$) domain. Here, $M = 0.25, t_2 = 0$. **d,** The band structure in the topological ($C = +1$) domain. Here, $M = 0.25, t_2 = 0.2$.

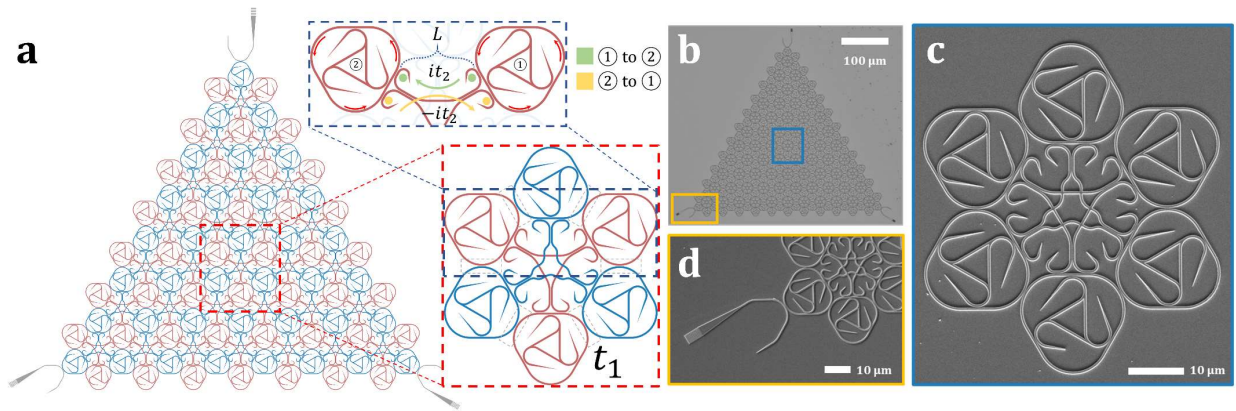

**Fig. 2 | The photonic topological Haldane lattice. a,** A schematic of an optical topological Haldane lattice with zigzag edges. Gratings with bus waveguides are placed at the three corners to monitor the output intensities. The inset in (**a**) depicts the Haldane honeycomb unit cell. Two sets of resonators having different perimeters (color coded by red and blue) are placed in close proximity to provide a NN coupling coefficient of $t_1$. The links are designed to normally intersect each other to minimize the crosstalk. The upper inset shows the required linkage for implementing the complex antisymmetric hopping between NNN resonators. **b,** A microscope image of the fabricated Haldane lattice on an InP wafer. **c,** An SEM image of a single Haldane honeycomb unit cell. **d,** An SEM image of the grating coupler at one corner of the Haldane lattice.

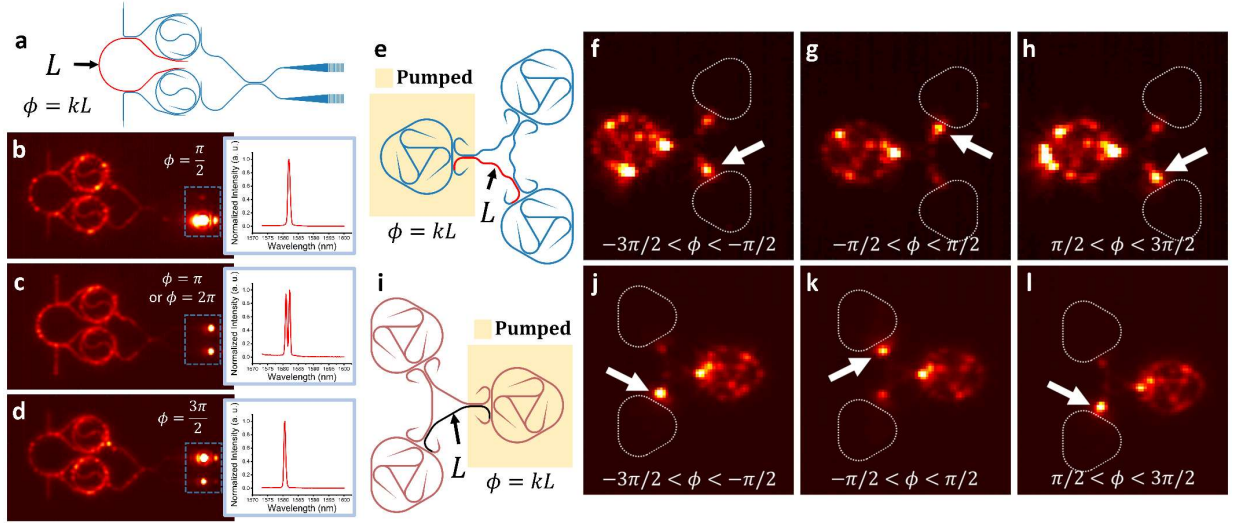

**Fig. 3 | Experimental results for 2-element and 3-element sub-units. a,** Schematic design of the 2-element system. Two unidirectional microring resonators with S-bends are coupled by a link. The phase $\phi$ is determined by the length of the link. Two gratings with a 3-dB coupler for interferometry is implemented as output. **b-d,** Experimental results of the intensity distributions for various values of $\phi$. Insets show the spectra at the output gratings. When the desired phase ($\phi = \pi$ or $\phi = 2\pi$) is achieved, the spectrum indicates that there are two lasing modes. **e, i,** Schematic design of the 3-element systems used for sub-lattice A and sub-lattice B, respectively. The phase $\phi$ is determined by the length of link $L$. In the experiments only one of the three resonators is pumped. **f-h, j-l,** The direction of the power flow towards the unpumped resonators changes as the phase $\phi$, varies from $-\pi$ to $\pi$.

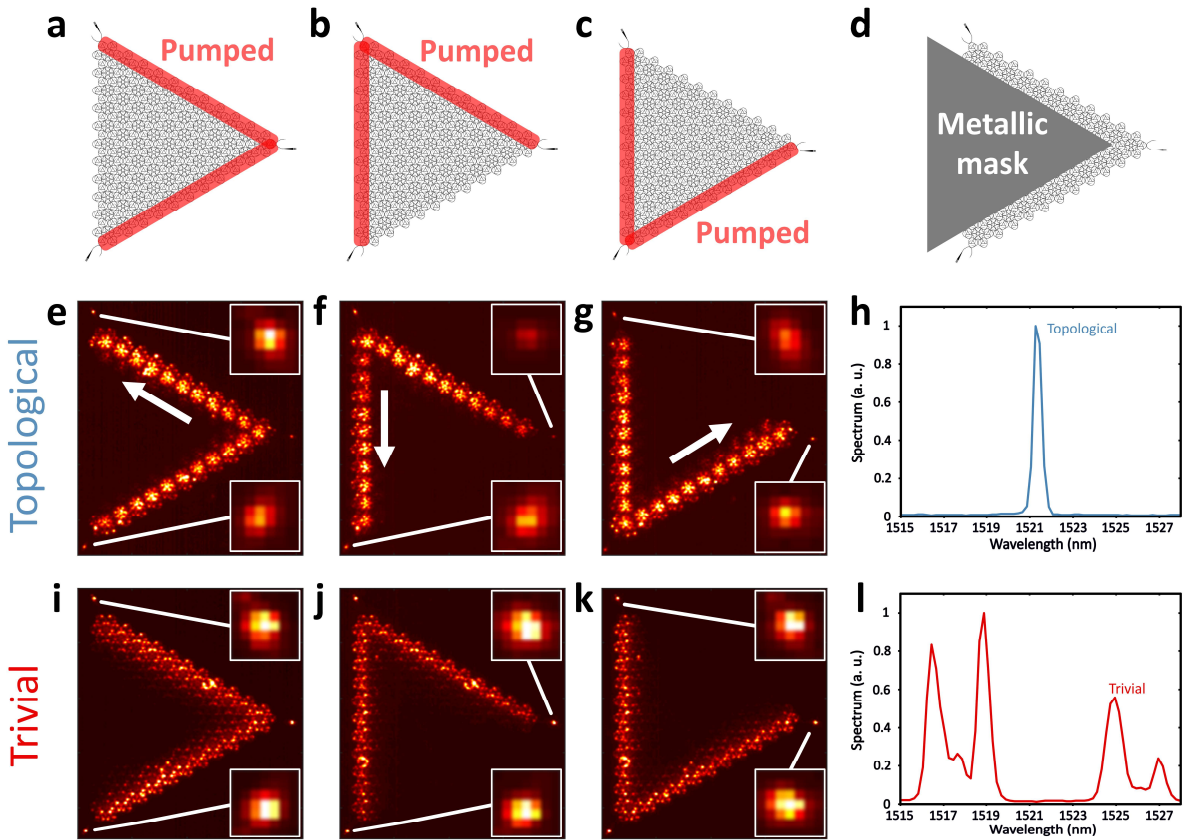

**Fig. 4 | Topological Haldane lattice - experimental results. a-c,** Optical beam patterns used to pump the Haldane lattice where in each experiment only two of the three edges are pumped. **d,** A triangular metallic mask is used to provide the required pump pattern. **e-g,** Intensity profile of the radiation emitted by this topological lattice for each of these three cases. The output intensities at the grating couplers are compared, indicating a counter-clockwise transport in the topological phase. The power ratio between the bright and the dark output intensities is 2.05, 3.68 and 2.03, respectively. **i-k,** Intensity profiles corresponding to the trivial Haldane lattice under the same pumping conditions. In this case, the intensities at the output gratings couplers are equal. The power ratio between the two output intensities is 1.02, 1.26 and 1.18, respectively. **h, l,** Emission spectra from the topological and trivial Haldane lattice, respectively. Note that due to highly suppressed back-reflections, the topological Haldane lattice exhibits a more uniform intensity distribution in its edge state as compared to the trivial counterpart.